\documentclass[sigconf]{acmart}

\usepackage{graphics}
\usepackage{colortbl}
\usepackage{cancel}
\AtBeginDocument{%
 \providecommand\BibTeX{{%
  \normalfont B\kern-0.5em{\scshape i\kern-0.25em b}\kern-0.8em\TeX}}}

\newcommand{\highlight}[1]{\textcolor{black}{#1}}
\newcommand{\cameraready}[1]{\textcolor{black}{#1}}

\copyrightyear{2024} 
\acmYear{2024} 
\setcopyright{acmlicensed}\acmConference[CHI '24]{Proceedings of the CHI Conference on Human Factors in Computing Systems}{May 11--16, 2024}{Honolulu, HI, USA}
\acmBooktitle{Proceedings of the CHI Conference on Human Factors in Computing Systems (CHI '24), May 11--16, 2024, Honolulu, HI, USA}
\acmDOI{10.1145/3613904.3641929}
\acmISBN{979-8-4007-0330-0/24/05}

\begin{document}





\title[\textit{Looking Together} $\neq$ \textit{Seeing the Same Thing}]{\textit{Looking Together} $\neq$ \textit{Seeing the Same Thing}: Understanding Surgeons' Visual Needs During Intra-operative Coordination and Instruction}

\author{Xinyue Chen}
\email{xinyuech@umich.edu}
\affiliation{%
  \institution{University of Michigan}
  \city{Ann Arbor}
  \state{Michigan}
  \country{USA}
}
\authornote{Both authors contributed equally to this work.}

\author{Vitaliy Popov}
\email{vipopov@umich.edu}
\affiliation{%
  \institution{University of Michigan}
  \city{Ann Arbor}
  \state{Michigan}
  \country{USA}
}
\authornotemark[1]

\author{Jingying Wang}
\email{wangchy@umich.edu}
\affiliation{%
  \institution{University of Michigan}
  \city{Ann Arbor}
  \state{Michigan}
  \country{USA}
}

\author{Michael Kemp}
\email{mikemp@med.umich.edu}
\affiliation{%
  \institution{Michigan Medicine}
  \city{Ann Arbor}
  \state{Michigan}
  \country{USA}
}

\author{Gurjit Sandhu}
\email{gurjit@med.umich.edu}
\affiliation{%
  \institution{Michigan Medicine}
  \city{Ann Arbor}
  \state{Michigan}
  \country{USA}
}

\author{Taylor Kantor}
\email{tkantor@med.umich.edu}
\affiliation{%
  \institution{Michigan Medicine}
  \city{Ann Arbor}
  \state{Michigan}
  \country{USA}
}

\author{Natalie Mateju}
\email{nmateju@umich.edu}
\affiliation{%
  \institution{University of Michigan}
  \city{Ann Arbor}
  \state{Michigan}
  \country{USA}
}

\author{Xu Wang}
\email{xwanghci@umich.edu}
\affiliation{%
  \institution{University of Michigan}
  \city{Ann Arbor}
  \state{Michigan}
  \country{USA}
}

\begin{abstract}

Shared gaze visualizations have been found to enhance collaboration and communication outcomes in diverse HCI scenarios including computer supported collaborative work and learning contexts. Given the importance of gaze in surgery operations, especially when a surgeon trainer and trainee need to coordinate their actions, research on the use of gaze to facilitate intra-operative coordination and instruction has been limited and shows mixed implications. We performed a field observation of 8 surgeries and an interview study with 14 surgeons to understand their visual needs during operations, informing ways to leverage and augment gaze to enhance intra-operative coordination and instruction.  We found that trainees have varying needs in receiving visual guidance which are often unfulfilled by the trainers’ instructions. It is critical for surgeons to control the timing of the gaze-based visualizations and effectively interpret gaze data. We suggest overlay technologies, e.g., gaze-based summaries and depth sensing, to augment raw gaze in support of surgical coordination and instruction. 
\end{abstract}




\maketitle

\section{Introduction}

One's gaze is a critical indicator of their visual attention \cite{zohary2022gaze}, and plays an important role in collaboration. \highlight{Gaze awareness is critical for the formation of joint visual attention (JVA), in which social partners focus on a common reference and monitor each other's gaze on external objects, people, or events \cite{van2016towards}.} It has been demonstrated that sharing gaze information facilitates, mediates, or regulates interaction between people involved in cooperative work and other forms of social activities. For example, sharing the collaborating partners' gaze enhances audio and text-based remote collaboration \cite{jing2022impact, kutt2020effects, lankes2022gazecues, vertegaal1999gaze} as well as in-person collaboration \cite{huang2019identifying, schneider2017real}, visualizing experts' gaze on videos increases novices' learning outcomes\cite{matsuda2021surgical}, sharing teachers' and students' gaze with each other benefits remote teaching and learning of physical tasks \cite{sung2021learners}.

Visual attention and gaze are critical for surgeons to successfully perform surgeries, especially in procedures where residents (trainees) collaborate with attending surgeons (trainers) in the operation room \footnote{https://www.aha.org/advocacy/teaching-hospitals} \cite{glarner2017resident}. Such intraoperative procedures \highlight{(the intraoperative phase begins when the patient enters the operating room
 and ends when the operation is complete)} pose a unique context to study the relationship between gaze and collaboration since the surgeons have dual purposes. On the one hand, the attending-resident dyad needs to coordinate to ensure patient safety; on the other hand, every procedure they perform together is a critical and precious learning opportunity for residents to gain surgical skills \cite{belyansky2011poor}. Prior work has demonstrated the importance of visual alignment for intra-operative coordination. For example, 97\% of surgical errors in the most common laparoscopic operation (laparoscopic cholecystectomy, \highlight{which is a minimally invasive surgical procedure to remove a diseased gallbladder}) resulted from visual misperceptions \cite{way2003causes}. Frequently used methods aimed at guiding visual attention to a target, including physical hand gestures and verbal explanations can easily be misinterpreted leading to increased risks to the patient's safety \cite{mentis2014learning, wilson2011gaze}. 92\% of the most senior residents report deficits in preparation for independent practice \cite{anderson2021defining}. New practices and understanding are therefore needed to help attending-resident dyads reach higher levels of visual alignment, and further help surgeons achieve both their operation, and teaching and learning goals.

Studies around the use and sharing of gaze in authentic operation room contexts are limited in contrast with other collaborative contexts. A notable example of providing visual guidance to enhance intraoperative coordination and instruction was performed in a simulated telementoring \highlight{(i.e., when an expert guides a novice medical provider remotely)} setting, where attending surgeons used telepointers \highlight{(i.e., a cursor that marks a point on a screen display)} to guide residents remotely \cite{semsar2019}. The study found that surgical trainees have challenges taking up and acting upon the information conveyed through telepointers, for reasons including trainees' tendency of relying on the telepointer instruction while ignoring other sources of information and insufficiency of the information provided by the telepointer. Other studies have shown that when only the trainers control the sharing of visual information, it impedes the learning of trainees since they can only passively accept the information instead of actively acquiring the knowledge \cite{feng2019communication, yuviler2011learning}. Such studies pointed out the challenges around sharing gaze and other visual cues during intraoperative procedures to support both operation, and teaching and learning at the same time.

In this work, we aim to address this important problem in health, how might we help the attending-resident surgeon dyads to visually align at critical moments when coordination and joint focus are needed to achieve their operation goals and optimize teaching and learning opportunities during surgery? Inspired by the use of gaze and visual guidance in CSCW and HCI literature, we aim to understand how residents and attending surgeons leverage visual information during an operation, and what visual information may be desired by residents and attending surgeons to enhance communication and coordination. This study aims to offer insights into how to provide visual and gaze-based support for surgeons to effectively coordinate, teach and learn in the operation room. 
To this end, we perform an interview and observation study. Specifically, we explore the following research questions:
\begin{itemize}
\item RQ1: What are the resident and attending surgeons' \highlight{visual needs and challenges around teaching and learning during laparoscopic surgery?}

\item RQ2: What challenges do resident and attending surgeons \highlight{face in coordinating and achieving visual alignment during laparoscopic surgery?}

\item RQ3: \highlight{How can we design practices to enhance the visual alignment between resident and attending surgeons, thus improving learning outcomes and ensuring patient safety?}

\end{itemize}

The interview and observation study involving 14 surgeons and \highlight{eight} authentic laparoscopic surgery cases show that resident surgeons' learning needs during operations are often unfulfilled. Specifically, junior residents need support identifying anatomical targets, \highlight{and surgical planes of dissection} and locations. Middle-level and senior residents want to operate as much as possible in a procedure. Attending surgeons at the same time also experience challenges \highlight{in} providing visual guidance to the residents. Residents see high utility in viewing the attending surgeons' gaze during critical portions of the surgery, while do not want attending surgeons to assess their competency based on gaze. Attending surgeons consider gaze to be a useful tool for teaching, and consider surgical competency to contain a lot more elements than visual alignment. Both residents and attending surgeons strongly support the use of gaze data for post-operative reflection and learning.

\section{Context}
This study focuses on one type of surgery, laparoscopic cholecystectomy, which is commonly known as gallbladder removal surgery. We will refer to the surgery as \textit{Lap Chole} in the rest of the paper. There are two reasons for situating our study in Lap Chole. First, it is the most basic surgery that every surgical resident needs to learn and master. 
Second, as a minimally invasive surgery, surgeons are guided by a video feed typically shown on two screens, as shown in Figure~\ref{fig:or}. On the one hand, the use of a screen makes the capturing and sharing of gaze information more tractable. On the other hand, visual alignment is in particular critical and challenging because surgeons are using a 2D screen to guide their 3D actions. 

\section{Related Work}

We first review prior literature on the use of gaze to support collaboration, the challenges on coordination and instruction in the operation room, and techniques to offer visual guidance in laparoscopic surgeries.

\subsection{Challenges on Coordination and Instruction in the Operation Room}

Current surgical training methods in the Operation Room (OR) pose challenges for both attending and resident surgeons in coordination, communication, instruction and learning \cite{matsuda2021, torres2022, aveling2018, belyansky2011poor}. For example, attending surgeons reported challenges in clearly articulating complex tacit skills and knowledge during a surgery, with one study finding they often utilize vague directives like "do some of this and some of that" when instructing residents \cite{matsuda2021}. Meanwhile, resident surgeons, conscious of the power dynamic and high-stakes environment in the OR, frequently struggle to voice concerns or disagreements regarding surgical decisions, even when they may have critical safety implications \cite{belyansky2011poor, coats2002, aveling2018}. For laparoscopic surgeries, limited depth perception \cite{bogdanova2016depth}, the loss of spatial orientation \cite{keehner2004spatial, sandor2010, wynsberghe2008}, and the need to interpret a 2D view of a 3D operative field \cite{sidhu2004interpretation, pierorazio2009} make these type of surgeries more difficult and can increase the learning curve of inexperienced trainees \cite{cope2015making}. As examples, surgical tools may rotate in different directions in reality versus on screen (camera view) \cite{wynsberghe2008, sandor2010minimally}. Surgeons' vision is scaled and bounded which poses challenges for trainees to relate what they see to the anatomy \cite{mentis2014learning}. New practices and understanding are therefore needed to help attending-resident dyads reach higher levels of visual alignment.

\subsection{Sharing Gaze in CSCW and CSCL Contexts}

Sharing gaze information has been found to facilitate collaboration, communication, and instruction across diverse Computer Supported \highlight{Collaborative} Work (CSCW) and Computer Supported Collaborative Learning (CSCL) contexts. Numerous studies have shown that visualizing gaze behaviors to collaborators helps them quickly understand each other's shared interests and leads to enhanced communication and collaboration outcomes. For example, Jing et al. \cite{jing2022impact} showed that in a remote collaboration involving a physical task, gaze visualizations amplified meaningful joint attention, and lowered collaborators' cognitive load while improving mutual understanding. K{\"u}tt et al. showed that sharing the collaborators' gaze helped people talk more about the shared content in both text-based and audio-based remote collaboration \cite{kutt2020effects}. Lankes and Gomez \cite{lankes2022gazecues} explored methods to visualize gaze cues in collaborative gaming environments, providing gaze cues through different shapes, e.g., spheres, lights, etc. 

Studies have shown the benefit of using gaze in teaching and learning scenarios as well. For example, visualizing experts' gaze to novices increases novices' learning outcomes across various domains including programming \cite{villamor2018predicting}, academic paper reading \cite{zhou2022does}, and video-based learning\cite{matsuda2021surgical}.
Sharing teachers' and students' gaze with each other benefits remote teaching and learning of physical tasks \cite{sung2021learners}. Schneider and Pea showed that real-time mutual gaze perception enhances collaborative learning outcomes and collaboration quality \cite{schneider2017real}.

What makes intraoperative procedures unique is that surgeons have dual purposes. On the one hand, the attending-resident dyad needs to coordinate to ensure patient safety during complex procedures that require high levels of visuospatial cognition and technical skill. On the other hand, every procedure they perform together is a critical and precious learning opportunity for residents to gain the advanced surgical skills and tacit knowledge that come from direct experience in the OR under expert supervision \cite{belyansky2011poor}. Balancing these dual goals of safe operative outcomes and effective on-the-job training places unique demands on the attending-resident team \cite{ferrier2023learning}. In this work, we study surgeons' needs and challenges during surgery procedures with the goal of providing gaze-based support to enhance intraoperative coordination and instruction at the same time.

\subsection{Visual Guidance in Laparoscopic Surgeries}
Given the importance of visual attention in laparoscopic surgeries, there has been significant interest in both the HCI and surgical education communities to explore gaze-based guidance and training techniques. Prior work can be summarized into three threads. First, enabling attending surgeons to use telepointers to provide real-time instruction; \highlight{s}econd, analyzing novice and expert surgeons' gaze behaviors posthoc and giving people feedback; \highlight{t}hird, simulation and virtual reality-based training systems that show trainees annotated content. 

First, researchers have had success using experts' gaze-overlay to improve trainees' performance in simulated settings \cite{chetwood2012collaborative, ashraf2018eye, feng2020virtual, rambourg2016}. However, recent studies also revealed drawbacks of offering single-user gaze information in simulated telementoring settings. Semsar et al. \cite{semsar2019} found that surgical trainees have challenges taking up and acting upon the information conveyed through telepointers, for reasons including trainees' tendency of relying on the telepointer instruction while ignoring other sources of information and insufficiency of the information provided by the telepointer. Other studies have shown that in surgical contexts when only the trainers control the sharing of visual information, it impedes the learning of trainees since they can only passively accept the information instead of actively acquiring the knowledge \cite{feng2019communication, yuviler2011learning}. 

Second, prior work has shown success in differentiating the gaze patterns between surgical experts and novices \cite {ashraf2018eye, aresta2016state, chetwood2012collaborative}. For example, expert surgeons demonstrated substantially greater fixation of relevant anatomic targets in laparoscopic procedures \cite{wilson2010gaze}, which \highlight{is} associated with higher technical proficiency \cite{evans2020comparison}. Richstone et al. \cite{richstone2010eye} were able to reliably differentiate inexperienced and expert surgeons with classifiers built on eye-tracking data. 

Third, surgical training systems are developed th\highlight{r}ough mostly simulated environments, which enable surgeons to provide visual overlays beyond gaze. For example, experts can provide instruction through 3D annotations \cite{weibel2020} and display 3D model overlays in-game and Virtual Reality (VR) environments \cite{marks2007, kalavakonda2015, nikodem2015, christensen2018, zhou2022}.

\textbf{As a summary,} attending and resident surgeons face significant challenges in intraoperative coordination and instruction. Many such challenges result from visual misperception and miscommunication. Gaze-based visualizations have been found across diverse CSCW and CSCL contexts to enhance communication and collaboration outcomes. However, prior studies on providing visual guidance in surgeries pointed out that trainees found the information conveyed through telepointers to be insufficient and that existing telepointers and gaze-based techniques caused trainees to passively accept the information provided to them instead of actively acquiring the knowledge and skills. They want to understand the instructor's decision-making process and the rationale behind each step. Visual cues alone fail to provide this. To support intraoperative coordination and instruction, it is imperative to develop a better understanding of surgeons' visual challenges and needs during the procedure.

\section{Method}

To address our research questions, we performed an interview study, followed by an observation study with authentic Lap Chole surgeries in the OR. The study aims to serve two goals, 1) validating 
 and probing into surgeons' needs on receiving visual support, 2) understanding how to provide visual support to surgeons, e.g., through offering shared gaze visualizations, which has been found to be effective in other collaboration contexts.

\subsection{Participant Recruitment}
We carried out this IRB-approved research at a university teaching hospital with a large surgical volume in the United States. Since surgeons are often on shifts and have busy schedules, especially during COVID, it was very challenging to recruit surgeon participants. We sent recruitment advertisements to resident surgeons' mailing lists and recruited through \highlight{our network}. Fourteen surgeons participated in our study, with \highlight{seven} attending surgeons (Table~\ref{attending}) and \highlight{seven} resident surgeons (Table~\ref{resident}). \highlight{PGY stands for postgraduate year. Surgical residencies can last from three to seven years in the US.}

\begin{table}[h]
\begin{tabular}{cccc}
\hline
ID & Gender & Age                                 & Training Level \\ \hline
R1 & Male   & 21-30 years old  & PGY2           \\
R2 & Male   & 21-30 years old  & PGY4           \\
R3 & Female & 21-30 years old   & PGY4           \\
R4 & Male   & 31-40 years old  & PGY7           \\
R5 & Male   & 21-30 years old   & PGY2           \\
R6 & Female & 21-30 years old & PGY3           \\
R7 & Male   & 21-30 years old  & PGY3           \\ \hline
\end{tabular}
\caption{Demographic Information of the residents that participated in the interview study.}  
\label{resident}
\end{table}

\begin{table}[h]
\begin{tabular}{cccc}
\hline
ID & Gender & Age                         & Years of Attending \\ \hline
A1 & Female & 41-50 years old & 9                                \\
A2 & Male   & 41-50 years old & 10                               \\
A3 & Male   & 41-50 years old                      & 6                                \\
A4 & Female & 41-50 years old                      & 15                               \\
A5 & Male   & 41-50 years old                       & 10                               \\
A6 & Female & 31-40 years old                       & 5                                \\
A7 & Male   & 51-60 years old                        & 15                               \\ \hline
\end{tabular}
\caption{Demographic Information of the attending surgeons that participated in the interview study. The attending surgeons have a minimum of 5 years of experience with mentoring surgical trainees and extensive experience performing Lap Chole procedures.} 
\label{attending}
\end{table}

\subsection{Interview Procedure} 
\highlight{The 1:1 semi-structured interviews, conducted either with the resident or attending surgeon individually, lasted about one hour on Zoom and consisted of two main} portions focusing on: 1) understanding surgeons' challenges around perceiving visual information during the surgeries in order to meet their instruction and operation goals. 2) probing into participants' attitudes towards capturing, sharing and interpreting eye gaze during surgeries. For the first portion, participants were asked to recall their most recent experience of a Lap Chole surgery, and share the visual cues they relied on for teaching/learning, problem-solving, and communication. The interview questions were the same for attending and resident surgeons with the particular questions tailored to the individual participant based on their role. For example, we asked attending surgeons to share how they gave residents instructions, \textit{``What kind of visual information (landmarks), if any, do you use to give residents instructions?''} We also asked residents how they received instructions, \textit{``What kind of visual information (landmarks), if any, does the attending surgeon use to give you instruction?''} The full list of interview questions are included in the supplementary materials. 

In the second part, participants were shown a demo video where attending and resident surgeons were both wearing eye-tracking glasses and their eye gaze was captured throughout the surgery. We used this as a design probe to investigate participants' thoughts around capturing, sharing and interpreting eye gaze during surgeries, in particular how they may envision using this information to support operation goals and learning goals, and what are the perceived risks with this approach.

\subsection{Observation Procedure}
\begin{figure}[h]
    \centering
    \includegraphics[width = 0.45\textwidth]{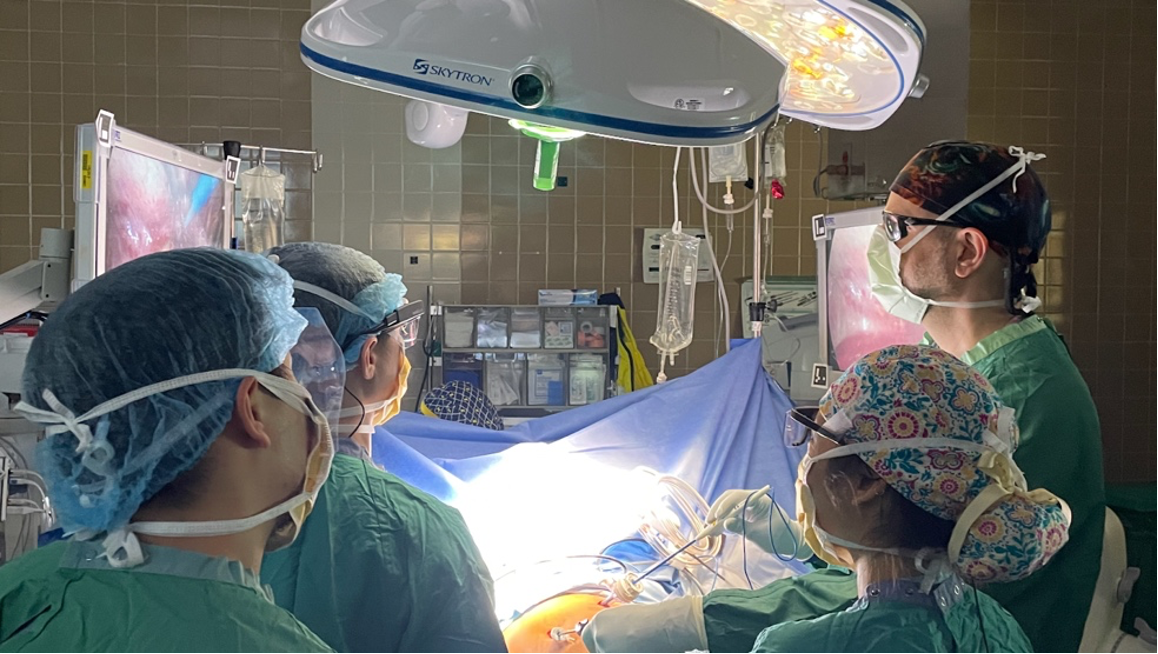}
    \caption{\highlight{A photo taken by the research team in one of the 8 surgeries we observed, which showed the data collection setup in the observations. Both attending and resident surgeons wore eye-tracking glasses, Tobii Pro Eye Glasses 2 with audio recording capabilities. Using eye-tracking glasses in the observations serves two goals. First, it allows us to record multi-modal data, including camera views from the eye-trackers, and the conversations between the surgeons. This is critical for interpreting the observation data since with audio data alone, it is insufficient for the researchers to fully understand the context. Second, it serves as an internal feasibility check of collecting eye gaze data while surgeons are performing the operation. One of the lead authors was present in the operation through all 8 surgeries.}}
    \label{fig:or}
\end{figure}

Following the interview study, we performed an IRB-approved follow-up observation of \highlight{eight} Lap Chole surgeries in the operation room. The goal of this observation study was to further understand the ways surgeons used visual information and the challenges they faced in communicating visual information during surgeries. In order to observe procedures in the operation room, we first obtained consent from the responsible attending surgeon on the days of the procedure. Three attending surgeons, two of whom participated in the interview study, allowed us into their operation room. We observed \highlight{eight} procedures in total involving three attending surgeons and five residents. Two of the residents were also participants of the interview study. On the day of the operation, both attending and resident surgeons wore eye-tracking glasses, i.e. Tobii Pro Eye Glasses 2 \cite{tobii2016tobii} with audio recording capabilities, to record their eye movements and the conversations between them. Using eye-tracking glasses in the observations serves two goals. First, it allows us to record multi-modal data, including camera views from the eye-trackers, and the conversations between the surgeons. This is critical for interpreting the observation data since with audio data alone, it is insufficient for the researchers to fully understand the context. Second, it serves as an internal feasibility check of collecting eye gaze data while surgeons are performing the operation. 
One of the lead authors was present in the operation through all \cameraready{eight} surgeries. 
In the operation room, the attending and resident surgeons usually stand on opposite sides of the patient and use two different monitors as shown in Figure~\ref{fig:or}, the content displayed on the two screens is the same. Figure~\ref{fig:or} is a photo taken by the research team in one of the \cameraready{eight} surgeries we observed.

\subsection{Data Analysis}

The interview recordings were first transcribed. We then conducted a thematic analysis \cite{braun2012thematic, braun2006using} using ATLAS.ti \footnote{https://atlasti.com}.
\cameraready{We took an inductive approach to construct themes in the data and focused on providing a rich thematic description of the dataset. Two authors first familiarized themselves with the data by reading through all transcripts and noting down initial ideas. They then generated initial codes inductively on two transcripts independently. The two authors met and went over each of their comments to discuss their interpretation. This process helped the lead author to familiarize with the data and triangulate their understanding of the data with another researcher. The lead author further continued generating initial codes for all of the transcripts through an interactive process, collating data relevant to each code. Themes were then constructed through an iterative process of reviewing and refining the codes to reflect broader patterns of meaning and relevant data were gathered under each theme. This involved revisiting the data multiple times, refining the themes to ensure they were coherent, and checking that they were well-supported by the data. The whole research team held weekly meetings to go over the codes and themes, ensuring that the finalized themes capture the key ideas in the dataset. }

\highlight{The two authors who did the initial coding performed the interviews and had contextual knowledge about the interviews. The two authors were also skilled in qualitative analysis. One of them had expert knowledge whereas the other had passing knowledge in surgery operations. Both authors had access to expert surgeons on the research team. We had monthly meetings with the surgeons to share progress and address questions. 
}

We analyzed the observation data to triangulate the findings from the interview \highlight{\cite{triangulation2014use}}.\highlight{The video and audio recordings of the surgeries were collected through the Tobii eye trackers the surgeons' wore during the operations. We used an AI transcription tool Otter.ai \footnote{https://otter.ai/} to generate an initial draft of the transcript based on the audio recordings, and the AI mistakes were further corrected by the authors. We coded the transcripts while replaying the recorded video. 
We used a similar thematic analysis approach \cite{braun2006using, braun2012thematic} to analyze the observation data. One author first familiarized herself with the observation data by reading through all the transcripts. She then did initial coding of all the transcripts through an iterative process. Since the observations are quite long, she particularly looked for moments when the residents demonstrated difficulties in understanding the instructions or finishing the procedure, and the moments involving any visual information. Then, the lead author constructed initial themes by examining the initial codes and collating relevant data. The themes were further developed through iterative analysis of the coded data and the observation dataset. The themes from the observation data were then merged with the themes generated from the interview data.
Finally, we grouped all the themes into three higher-level themes: ``residents' learning challenges that are not fulfilled by attending surgeons' instruction'', ``visual cues used in surgery and the perceived visual challenges'', ``potential utility and barriers of using Joint Visual Attention in surgical training''.}

\section{Findings}

Overall, compared to general CSCW, our findings show that surgical gaze sharing must consider residents' \highlight{skill level}, power imbalances, dual goals of efficiency and training, and patient safety risks \cite{ferrier2023learning}. The surgical education context requires tailored approaches to realize the benefits.

\subsection{Residents' challenges on receiving and presenting visual information during operation and their needs for support}

An important theme emerging from the data is around residents' learning experiences during the procedure. Residents shared frustrated moments where their learning needs were not fulfilled by the attending surgeons' instructions. Notably, residents cited different learning needs at different stages in their training, implying that a "one-size-fits-all" instructional approach may not apply.

\subsubsection{Junior residents need support identifying anatomical targets, planes and locations in a 3D space}

Most of the interviewed junior residents shared that they encountered challenges mapping anatomic planes \highlight{i.e., an imaginary flat surface that is used to define a particular area of anatomy,} or structures from textbooks, including blood vessels and organs, to the structures in a real human body in the operation room. R3 said, \textit{"For a first-year resident, their learning goal might be that they are able to successfully identify the planes because it is different from what you were looking at the textbook before, there’s like several different layers when you see inside the body."} 
Residents commented that the attending surgeons' verbal instructions could be vague and insufficient. R7 noted: \textit{"I definitely remember times when an attending will be explaining something, and the way they were explaining it just was not helpful. And I think the area where I struggled the most was achieving the appropriate plane between the gallbladder and liver. And I think that's more of an error of like visual familiarity and visually identifying something that hasn't been verbally taught well by the attending."} 
Residents further explained that the attending surgeons' explanations were in particular insufficient when they needed to identify angles and directions. R2 said, \textit{"The surgeon might just say he is taking stuff over there. And I don’t think that’s very helpful for me because I did not know what they were talking about."} 
Residents mentioned that the visual misconceptions could be subtle while leading to serious consequences. R3 said, \textit{``Sometimes I thought I saw that but what I actually saw was to go more to the left instead of there. So it's often a more minor miscommunication, just a couple of centimeters. But it may lead to a huge error.''}

Echoing residents' experiences, in the observations, we found it common for attending surgeons to verbally describe spatial positions or refer to anatomical targets verbally, lacking precision and clarity. For example in O1, the attending surgeon was explaining \textit{"So I would say to clear this off, I'm gonna zip it off the hepatic class a little bit and then clean it up."}
Another example in O6 with frequent use of words to describe orientation, \textit{ ``
Okay, now you're going to just open up some of that on the left side of the duct. Don't touch the deck to the left, all the way over here. It's higher now. A little bit down. Right there.."}

\subsubsection{Junior residents can't perceive the same visual information on the screen as attending surgeons in order to make procedural decisions}
One of the most important skills residents are developing in the operation room is decision making. Surgeons need to review the patient's anatomy and the current progress in order to make plans for the next steps. Junior residents shared that it was challenging for them to perceive the same visual information as attending surgeons, making it difficult for them to know what was the basis of the decisions made by the attending surgeons. This happens more frequently with junior residents when they observe attending surgeons operate. 
R5 said, \textit{"If you've only watched someone do the operation, he saw the struggle view \highlight{[i.e., a challenging moment in the surgery]} but he may forget to show it verbally and describe why what they're doing is very challenging. Then you may still don't know how to do it in the next week."}
R4 also described a similar situation where it was hard for them to effectively acquire knowledge since they missed the bigger picture, \textit{"Sometimes you’ll see something that just looks challenging, sounds challenging, and you’ve never done quite like that before. So it is challenging to define a proper learning goal.”}

\subsubsection{Middle-level and Senior Residents want to operate as much as possible and do not want the attending surgeons to take over}

Middle-level and senior residents shared that they wished to perform critical portions of the operation independently and safely. From the attending surgeon's perspective, they will continually monitor the resident’s ability to perform the procedure and will take over the procedure when they think it is unsafe. R2 said \textit{"To try and get more autonomy from the attending requires you to really know the step that you’re on and multiple steps moving forward. Otherwise, they’re going to take over from you. And there’s a worry that they’ll just keep going and you won't get to practice your part of the surgery."} 

Residents shared their strong desires to do as much of the operation as possible and that take-overs by attending surgeons were viewed by them as "failures". R4 said, \textit{“My goal is to do all of it. So do the entire procedure with guidance from the attending.”}
R5 provided an example, \textit{"I can remember one time very distinctly where attending became quite frustrated with me because he handed me a sort of instrument that I never used before. I began to use it in a certain way. I remember the first thing he said was 'no, you’re missing'. And then I kept trying. And he said, ‘did you really have the ability to move things in space?'. And he eventually took it from me."} 
Such take-overs place heavy pressure on the residents and may lead to emotional breakdowns. R6 said, \textit{"Sometimes when you get a surgery taken away from you, as a resident, it feels like you’ve messed up or you felt like something went wrong, you’re not good enough."}

Residents also explained \highlight{surgical} cases where they did not want to ask questions that they thought would undermine the trust between themselves and the attending surgeon. This is not ideal and may lead to errors. R2 said, \textit{"I think a lot of the time, if I'm not a hundred percent sure what somebody is asking me to look at, I just grabbed the thing that I think they're talking about or, or point to it and then ask if it is what they're talking about. And I think that slows down the case. And you are not building trust because you're always asking questions."}

\subsubsection{Senior residents want ways to express themselves}

Junior and middle-level residents are mainly receiving instructions during the procedure. However, senior residents are often given the autonomy to decide on the specific strategy to be used. Senior residents shared cases where they disagreed with the attending surgeon on how to perform a step. As shown in O5, the attending surgeon asked the resident to explain a step they were doing, \textit{"You’re gonna put in your lateral port? So I don’t know why you want to do that. Is it because he has pancreatic tissue?"}  

In addition, senior residents have a desire to develop mentoring skills. In one of the cases we observed, the senior resident was operating, the attending surgeon was monitoring the progress, and a junior resident was in the room observing. In this case, the senior resident was learning how the attending surgeon explained things to the junior resident, e.g., \highlight{A5} \textit{"The reason why it's important is you can even see there’s a bile duct. So this is a risk."} Senior residents mentioned that they wanted to have opportunities to practice mentoring skills in the operation room.

\subsubsection{Residents desire more feedback on their performance}

Residents of all stages shared the desire to receive continuous feedback on their performance during critical portions of the operation, but this need is often not met. It can be difficult for attending surgeons to narrate each step and appraise the residents for every correct manipulation. Some residents shared that it could be helpful for the attending surgeons to give them summarized feedback after each step in the procedure. As R7 said, \textit{"I think it could be potentially helpful to have like just a sort of subjective grade by the attending for different certain types of steps, like are we taking too much tissue or too little tissue? Or is the needle oriented? How many times do you accidentally drop them?"}

\subsection{Attending surgeons' challenges on providing visual guidance and their needs for support}

\subsubsection{Attending surgeons find it difficult to know whether the resident is looking at the right place}
Most interviewed attending surgeons shared that guiding trainees' visual attention to a specific target was a major focus and challenge in laparoscopic training. Many attending surgeons shared similar sentiment that, \textit{"It's hard to know if the resident is looking at the right place."} In the operation room, as shown in Figure~\ref{fig:or}, the two surgeons stand on different sides of the patient's bed and view two separate screens. This setup further contributed to the difficulty in achieving visual alignment. For example, A3 said:\textit{"Sometimes instead of focusing on my screen, I turn around and focus on their screen to see whether what they're seeing and what I'm seeing is the same."} Surgeons further described scenarios that led to visual misperceptions due to this setup. As A3 said, \textit{"Let's say there are three bleeders. But the first one is more prominent on their screen. It does happen just because of the way the intensity angles are, and there could be visual disagreement on the information that’s being conveyed."} A6 said, \textit{"If you just say left, they're moving to the other side. So we need to make it very clear. I will say it's screen left, not patient left."}

\subsubsection{Attending surgeons find it hard to verbally communicate where the residents should be looking at.}
Attending surgeons are limited in their hand movements and other visual references to help their verbal explanations in surgery, especially when they work with junior residents during which the attending surgeons operate a majority of the procedure. A1 mentioned: \textit{"I don't have a way or an extra hand to point on the screen to show them."} They feel that communicating the visual information verbally itself is really difficult due to the different mental models and different experiences between the attending and residents, as A2 said, \textit{"Some things are really challenging to communicate because they are just experiential. And you either see it or you don't. No matter what has been explained, there is like, if you haven't experienced it, or seen it, it becomes very hard to do it."}

\subsection{\highlight{Attending and resident surgeons'} attitudes towards capturing, sharing and interpreting gaze during operations.}

\subsubsection{Residents see the high utility in viewing the attending surgeons' gaze during critical portions of the surgery}
Residents reported that watching attending surgeons’ in-the-moment gaze behavior could help them better understand why, when, how, and where to focus their attention. They thought real-time gaze could be especially helpful when they needed to locate a position. R3 said, \textit{"Sometimes there is very subtle differences in the position that may be desired. So I would imagine perhaps seeing exactly where they're looking and where they want you to make the nice incision. So maybe there could be some assistance, like, okay, you're looking here, and pressure at here."}   
Residents also commented that seeing the attending surgeon's gaze may give them confidence and reduce stress.

\subsubsection{Attending surgeons consider gaze a useful tool to teach}
The attending surgeons liked to be able to use gaze to instruct, similar to using telepointers. A2 said,\textit{"I describe things like, something has to be on stretch, but it can't be stretched to the point where it's going to break and it can't be too loose. And this is super hard to describe. Using attention as a visual cue can help."} 
The attending surgeons also commented that visualizing the residents' gaze could help them assess residents' performance and avoid safety issues. 
As A5 said, \textit{"I want to know his attention during the moment when he has to be absolutely in tandem with what I'm suggesting. It's in clipping and dividing those two structures, the cystic artery and the cystic duct. If he goes lateral, or if he is in the wrong plane, he can injure the duct. And then we'll be in trouble. I want to know their vision exactly."} Knowing the residents' visual attention may also free up the attending surgeons' cognitive capacity so that they can focus more on driving the operation safely and providing instruction to the residents. 
As A6 said, \textit{`` If I know someone's seeing something the way that I'm seeing it, then I'm more focused on what they're doing. I can focus more technically, then, competency-wise, you know. Without the gaze data,  I always in the back of my mind, I'm like, are we definitely looking at them? But with this, I can think more about whether is this the right thing. ''}

\subsubsection{Both residents and attending surgeons expressed concerns about using gaze to assess resident's competency}
Both residents and attending surgeons shared concerns about using the gaze alignment between themselves as an indicator of the residents' competency. First, the procedure can be long and there are many non-operational actions such as role-switching and interpersonal communication, attending and resident surgeons are not required to visually converge at all times. R4 said, \textit{"I think you can focus the analysis on the critical portions of the surgery, it's in general where the attending pays the most attention to during a procedure. And then looking at the same thing at that time is a fair analysis."}
A5 said \textit{"So in essence, how do visual connection and contact with a particular object suggest entrustment? If both of us are looking at the same object, that means there is a higher score for that entrustment? But sometimes he may be doing something, but I'm actually not looking at that part to judge a particular system with entrustment."} 

Second, surgical competency contains a lot more elements than visual alignment. Most attending surgeons agreed that visual alignment can be limited in predicting technical skills, efficiency and safety. A2 noted \textit{"if you're both looking at the same thing, but I don't know that it's a proxy of how well their hands are, just because they're looking at the right thing...e.g., whether their tissue handling is appropriate, whether the utilization of the instrument is appropriate, whether they know the actual steps of the operation. So there's like a fair number of other variables that can impact the treatment."}
Attending surgeons instead proposed to use gaze alignment as an alert for the surgeons to reflect on their cognitive alignment beyond gaze. A6 said,  \textit{"Now as a human mentor, if you notice somebody does it three or four times, you kind of view that's the signal to you. And then gaze alignment becomes an extra element in the operation. For example, maybe when the resident is struggling, the mismatch of gaze could be an extra indicator to tell you that.''}

Third, residents are concerned about false alerts that may lead to their operation opportunity being taken away unnecessarily. Alternatively, residents recommended using gaze alignment to indicate the attending surgeons' communication efficacy. For example, R5 said, \textit{"That's why it can be helpful to the communicator as a tool to say, Hey look, your communication, you got as a minus because you did it pretty well. But miss this, or you got a D minus, because you told them to go look this way and they kept looking over here."}

\subsection{\highlight{Both attending and resident surgeons} expressed strong support to use gaze information post an operation for reflection and learning}

There was a clear sense among the attending surgeons that the gaze behaviors of the residents and the contrast of gaze between the dyad, could help identify teachable moments after the operation is over. 
Attending surgeons considered this could increase the efficiency on offering feedback during the debrief of a procedure. 
A5 noted \textit{“there's a way to provide that data back and say hey, at minute 10 it was off, at minute 14 and it was off at minute 22 it was off and then focus on those moments where it was different and then start the conversation with coaching and provide feedback on those moments."} 
Attending surgeons expressed interest to review and interpret gaze data if such data can be structured and visualized for them. A4 said they would look for data anomalies, e.g., \textit{"If  here's a snippet of 30 seconds where you were the most discordant. And we can look at it. That's useful, like, Okay, why were we discordant there?"} 
Residents shared similar desire to review gaze alignment with the attending surgeon afterwards. For example, R5 said \textit{"And then in the debrief afterwards, when you see this scenario, you should think about this. you know, the anatomy can look this particular way and here's the way you want to approach it."} Similarly, R4 said \textit{"But I could see there's some value in it being like debrief tool after the fact where you can say, Oh, I was looking over at this particular structure while you were looking here."}

Attending surgeons shared that they could review the videos and gaze distance between the dyad to understand when they presented information effectively. They can thus work on improving their communication and teaching strategies.
As A6 noted \textit{"What kind of cues make the resident’s eyes go from right field to where I [attending] want them to look...There are clear implications in relation to communication and cognitive load, specifically in the operating room. This is in face using technology to evaluate how we communicate in the operating room."}

\section{Discussion and Design Implications}

This research explores the challenges and potential of gaze sharing in the unique context of surgical training, contributing new insights to the broader HCI and CSCW discourse on the use of gaze in high-stakes collaborative settings. Our study highlights the evolving learning needs of surgical residents at different stages of training and the limitations of current instructional approaches in fulfilling these needs. Surgeons recognized the value of gaze data in enhancing learning and communication during surgical training. However, surgeons also expressed concerns about gaze data misinterpretation and the need for control over its visualization to avoid distraction. Our findings underscore the complexity and distinctive aspects of integrating CSCW and CSCL strategies into surgical contexts. Among these are facilitating resident development of their professional vision, navigating power differentials, balancing efficiency and training, and ensuring patient safety. This study also provides design implications associated with sharing gaze data in a technically advanced and constrained setting. While this study made valuable contributions, there are limitations to acknowledge. First, we performed observations of only eight surgeries, a relatively small sample. Second, the observations were situated within a specific cultural context of a large US teaching hospital, constraining potential generlizability of the findings to other (international and/or low resource) training environments which may involve different OR team dynamics, norms, and educational approaches.

\subsection{Contextual visual guidance and feedback is needed in intraoperative coordination and instruction}

Our study suggested that both attending and resident surgeons experienced substantial visual challenges during operations. The challenges are exacerbated by the fact that surgeons have both operational and educational goals. As examples, junior residents cannot perceive the same visual information on the screen as attending surgeons, which suggests that simply guiding residents' attention to a particular scene is not sufficient for learning. More contextual information needs to be provided for residents to fully understand what visual cues the attending surgeon is leveraging to make operational decisions. Moreover, it is evident from the interviews and observations that the attending surgeon's instruction and feedback provision could be enhanced by richer visual information. Residents cite a lack of feedback mechanisms for them to effectively learn and gain independence. For example, junior residents may want more feedback on their performance, whereas senior residents may want more feedback and channels to elicit feedback from attending surgeons in order to independently operate. From the attending surgeons' perspective, they also experience difficulty verbally communicating where the residents should be looking at and "assessing" the residents' state of mind in order to give them feedback. Our study provides rich data on the scenarios where attending and resident surgeons cannot achieve their operation, teaching and learning goals when visual information, including surgeons' attention information is lacking. Our study provides strong motivation for designs where residents can receive more visual feedback from the attending surgeons and have channels to elicit feedback from the attending surgeons, and where the attending surgeons have effective methods to guide residents' attention, and provide instruction and feedback. The scenarios we present in this paper align with prior work \cite{semsar2019, feng2019communication, yuviler2011learning} in suggesting that telepointers are insufficient. On the one hand they are attending surgeon focused and do not represent residents' state of mind. On the other hand, they do not contain sufficient information for residents to learn. Our study further explores the design space of providing shared gaze visualizations between the surgical dyad. We will discuss in the next section the design requirements derived from this study.

\subsection{Gaze data is viewed as an anchor for attending and resident surgeons to analyze the situations and identify opportunities for instruction and remediation.}
A unique property of gaze-sharing in surgery context is that there is 
pronounced power difference between attending surgeons and trainees \cite{hu2012protecting, hu2016surgeons}. Junior residents in our study reported being reluctant to ask clarification questions when instructions were unclear, yet they needed precise spatial guidance from attending surgeons. Senior residents wanted channels to suggest alternative surgical approaches and techniques. These tensions align with literature emphasizing the need to open bi-directional communication channels and increase transparency in high-power distance collaborations \cite{hinds2015flow}. Surgeons view the use of gaze as an anchor for them to provide further explanation and analysis of the situations and opportunities for remediation. Surgeons do not consider gaze alignment to necessarily indicate competency but consider mismatches to be a signal of risks. Gaze sharing could provide pathways for residents to express needs and give attending surgeons insights into the residents' visual understanding. For example, detected gaze discrepancies might cue attending surgeons that additional explanation is required without requiring residents to explicitly request help. This is consistent with prior work showing that giving instructions on moving the cameras during surgery requires everyone to constantly coordinate based on what is happening \cite{mondada2016operating}. Additionally, our study aligns with prior work in revealing that residents have a wide range of expertise \cite{cheatle2019sensing, mentis2014learning}, where attending surgeons' instructions need to adapt to residents' prior knowledge in order to achieve optimal teaching and learning outcomes. Residents' gaze may provide useful information for attending surgeons' to assess their state of mind and background knowledge, before offering more adaptive guidance \cite{cheatle2019sensing}. Our study presents unique attributes of gaze-sharing in intraoperative contexts where there is power imbalance, the task is highly visual and complex, and the participating partners have dual purposes of achieving operation and educational goals.

\subsection{Implications for designing gaze-based technologies to enhance intraoperative coordination and instruction}

\subsubsection{Intraoperative gaze-sharing should be on-demand}
Prior work on attention and distraction suggests user control can mitigate effects on cognitive load \cite{plopski2022eye}. Attending surgeons in our study echoed this need to control gaze sharing to avoid excessive distraction from their primary surgical tasks. Participants provided suggestions aligned with prior literature, including on-demand gaze sharing triggered by foot pedals or voice commands rather than continuous display. A3 noted\textit{"... I think having it on all the time will be too distracting honestly. Stressful to know that your resident is not looking where you're asking them. You could have it selectively on, but interacting with technology while you're in a sterile field is difficult."} A3 further illustrated what he meant by on-demand: \textit{"A foot pedal would be nice. Like if I had a little foot pedal that I could like, tap it whenever I want to see where they're looking, might be useful."}  
Allowing \highlight{attending and resident surgeons} to initiate gaze sharing when situations warrant mitigates the risks brought up in prior literature that exclusive trainer-driven visual guidance can limit trainee agency and autonomy \cite{feng2019}. Resident surgeons preferred opportunities to interpret the meaning behind surgical moves and strategies, not just passive observation. Enabling trainees to request gaze sharing gives them more autonomy in leveraging this information source when needed. Overall, these findings imply gaze sharing systems should provide flexible user control over when visualizations appear rather than constant display.

\subsubsection{Gaze-based visualizations and AR overlays to enhance teaching and feedback during operations}

Besides simply overlaying the gaze patterns, surgeons want to get more support on looking beyond the "gaze alignment" data. Based on the scenarios we presented in this paper, we make design recommendations of gaze-based visualizations and AR overlays that may facilitate intraoperative coordination and instruction. First, there are critical anatomical components to the lap chole procedures, e.g., gallbladder, arteries, etc. Visualizing and tracking the region of the surgeons' gaze might be more productive than tracking the raw coordinates of the surgeons' gaze. Recent work on surgery scene segmentation techniques makes this direction more tractable \cite{wang2024surgment}. Second, our study revealed that novice surgeons encountered difficulty in perceiving depth in 2D images. Providing depth sensing scaffolds on surgical scenes could be helpful, e.g., visualizing the size of the gaze point to help people perceive spatial relationships. Third, we found that when attending surgeons explained their decision making process, it often involved the using of hand drawing in the air. Enabling surgeons to easily make visual annotations could support the instruction process. Recent work on surgery scene segmentation techniques \cite{wang2024surgment} may enable surgeons' to highlight anatomical structures in real time during intraoperative instruction. Fourth, we found it common for the attending surgeon to refer to a previous scene or a previous step when explaining the current step. AR-techniques that explore playbacks maybe desirable. Lastly, we consider it beneficial to study the use of gaze-based analytics in surgical contexts. For example, prior work has shown that the number of fixations per time interval, the mean fixation duration, and the index of pupillary activity are the three indicators correlated with the expertise level of the surgeons \cite{gunawardena2019, ashraf2018eye}. In addition, recent work by Abdou et al. \cite{abdou2022gaze} and O'Dwyer et al. \cite{o2019eye} summarized the commonly used gaze features that are successful in predicting emotion, particularly gaze location, gaze angle, pupil diameter, eye blink intensity, etc. Exploring visualization techniques to summarize gaze information and visualize them in real-time to facilitate instruction, reflection and feedback may be a fruitful direction. Moreover, such gaze-based analytics can be leveraged to generate safety alerts to surgeons when the alignment between the attending and the resident surgeon is below a threshold.

\subsubsection{Tradeoffs emerge between uni- and bi-directional gaze sharing}

Aligning with prior work \cite{semsar2019, feng2019communication, yuviler2011learning}, our study suggests that merely allowing attending surgeons to use telepointers is insufficient for intraoperative instruction and coordination. On the other hand, our work also suggests tradeoffs between uni- and bi-directional gaze sharing between the attending and resident surgeons. Both attenting and resident surgeons responded positively to bi-directional gaze sharing in enhancing transparency and supporting teaching and learning. However, both attending and resident surgeons mentioned scenarios where bi-directional gaze sharing is undesirable. For example, attending surgeons mentioned that for uncritical portions of the surgery, bi-directional gaze sharing may be unproductive and distracting. Residents want bi-directional gaze access to provide helpful grounding and context. Junior residents in particular reported this continuous awareness would aid their visuospatial understanding and "professional vision" development \cite{mentis2014learning, mentis2020}. However, residents also do not want their gaze data to be used as the only basis for assessment.

\subsection{Implications for designing technologies to enhance postoperative learning and reflection}

Providing joint visual attention data post-operatively, especially extracting data points where surgeons experience difficulty in aligning their visual attention, can provide significant insights during post-operative debriefing for attending surgeons to give residents feedback \cite{zheng2016team, matsuda2021}. As reported by our attending surgeon participants, evidence-based guidance to trainees is surprisingly limited, despite the importance of timely feedback for trainees to learn. Both resident and attending surgeons mentioned the need and desire to combine eye-tracking data and OR video recordings for trainees to reflect on their operational performance. \cameraready{To this end, future research could combine OR video recordings and eye-gaze data to help surgeons navigate a surgery recording \cite{avellino2021surgical} based on the attending and resident surgeons' misalignment.
Moreover, recent work has explored using surgery scene segmentation techniques \cite{wang2024surgment} to help surgeons semantically search critical scenes of surgery recordings. Such techniques make it possible to embrace the idea of "semantic gaze", where systems can infer the specific anatomical structures surgeons are looking at based on their raw gaze position on the screen. Future work may also explore the possibility of creating video-based tutorials with gaze overlays as training materials for residents \cite{grossman2010chronicle, matejka2013swifter}.}

\subsection{Ethical considerations on using gaze data in surgical training and collaboration}
As with any technology, we must also consider the ethical implications of eye-tracking technology, especially in a high-stakes environment such as the operating room \cite{prigoff2016ethical}. There are many potential risks associated with the use of eye-tracking data and gaze-based analytics, including the potential for biased decision-making, the loss of privacy, Hawthorne effects \cite{sedgwick2015understanding} (which refers to the phenomenon that individuals modify an aspect of their behavior in response to their awareness of being observed), and the potential for misuse \cite{bachynskyi2014motion, sadeghi2022systematic}. It is important to consider these risks carefully and to put in place measures to mitigate them. As mentioned by the attending surgeons, gaze data could only indicate a small portion of the resident's surgical competency. It requires careful work to examine what gaze alignment data can tell us and what could be missing in such data. For example, one may correlate residents' gaze performance with validated surgery skill assessment scales (e.g., the OpTrust, SIMPL \cite{sandhu2018optrust, zwisch}) to investigate whether gaze-based metrics is a reliable assessment metric for some elements in one's surgical competency.

\section{Conclusion}

This study makes two valuable contributions to the body of research on leveraging eye-tracking and gaze data to enhance surgical training. First, through interviews with 14 surgeons and observations of 8 authentic surgeries, this paper reveals key tensions around spotlighting visual attention in an environment characterized by varying trainee needs, power differentials, and dual priorities of patient safety and instruction. Second, this study demonstrates the potential benefits of capturing real-time gaze data from surgeons in operation rooms and provides design implications guiding the development of mixed-reality surgical environments to foster intra- and postoperative teaching and learning. There is tremendous potential for gaze-enhanced systems to facilitate resident "professional vision" development, open bi-directional communication channels, provide personalized support matched to skill level, and enable rich team reflexivity opportunities. With careful design informed by our study findings, sharing gaze data has promise for improving coordination, enhancing safety, fostering learning gains, and revealing invisible facets of collaborative cognition across surgical teams. Our findings help lay the groundwork for innovative technical systems that ultimately aim to improve surgical training using multi-modal, including gaze-based technologies.

\bibliographystyle{ACM-Reference-Format}
\bibliography{references}

\end{document}